# Chapter 6: An Introduction to RNA Databases

Marc P. Hoeppner [1], Lars E. Barquist [2] and Paul P. Gardner [2,3]

1 Department of Molecular Biology and Functional Genomics, Stockholm University, Stockholm, Sweden

2 Wellcome Trust Sanger Institute, Wellcome Trust Genome Campus, Hinxton CB10 1SA0, UK

3 School of Biological Sciences, University of Canterbury, Christchurch, New Zealand

Marc P. Hoeppner: marc.hoeppner@molbio.su.se
Lars E. Barquist: lb14@sanger.ac.uk
Paul P. Gardner: pg5@sanger.ac.uk

## *Abstract*

We present an introduction to RNA databases. The history and technology behind RNA databases is briefly discussed. We examine differing methods of data collection and curation, and discuss their impact on both the scope and accuracy of the resulting databases. Finally, we demonstrate these principals through detailed examination of four leading RNA databases: Noncode, miRBase, Rfam, and SILVA.

## *Introduction*

The introduction of targeted molecular and bioinformatic approaches (1,2) and the availability of affordable sequencing technologies have lead to a glut of novel ncRNA sequences (Figure 6.1). NcRNAs have been shown to be involved in a diverse array of cellular processes, from long-known roles in the translational process to more recently discovered functions in the regulation of gene expression and genomic defense (3,4). Databases provide a central resource for researchers to obtain and deposit this information in the form of sequences and descriptive metadata.

**[Figure 6.1]**

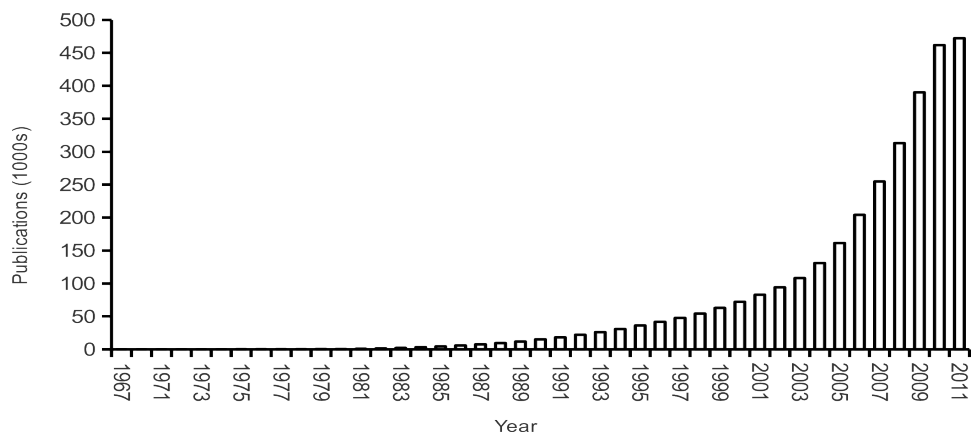

The purpose of this chapter is to introduce the reader to the state of the art of RNA databases. The first section of this chapter will provide a brief history of RNA databases which will put current databases in context. The second section covers approaches to data collection and curation, and how these differing approaches affect the utility of data served. Finally, we illustrate these points by exploring a few exemplar databases in depth: Noncode, miRBase, Rfam, and SILVA.

## *RNA databases - a historical perspective*

The development of modern RNA databases reflects both our growing knowledge of RNA biology and the development of modern information technologies. The earliest databases focused on long-known classes of RNA molecules. For instance, the original release of the Sprinzl tRNA database was distributed as the text of a journal article (5). Other early databases include the Signal Recognition Particle Database (6) and the Ribosomal RNA Database project (7), both distributed as "flat" text files over FTP.

The growing number of sequenced RNAs drove a shift in database technology. While the original Sprinzl tRNA database contained approximately 700 sequences, it's modern inheritor contains more than 12,000 tRNA genes (8), and the current release of the Rfam database contains over 1 million sequences computationally identified as encoding tRNAs. This amount of data would be impractical to store and search as flat files. The solution to this problem that has been adopted is the use of relational databases accessible over the world wide web.

The strength of relational databases lies in the ability to have one data repository from which multiple outputs can be generated dynamically. Any updates to the database will affect all output. In contrast, flat files need to be updated individually. More specifically, a relational database acts as a server which organizes data and provides it in an interactive fashion to client applications. All output draws from the same source, making data maintenance considerably easier. In addition, the way data is organized allows for the use of complex boolean queries to retrieve specific information of interest. An example would be collecting all human ncRNAs that belong to a specific class, are between 50 and 60 nucleotides long and were published after 2004. This would require query-specific scripting with flat file data, and would then require constant updating of a multitude of files as new information becomes available. In a relational database, we would simply select the relevant entries from our tables, and any changes in the underlying data set could be captured by re-running our query.

[Figure 6.2]

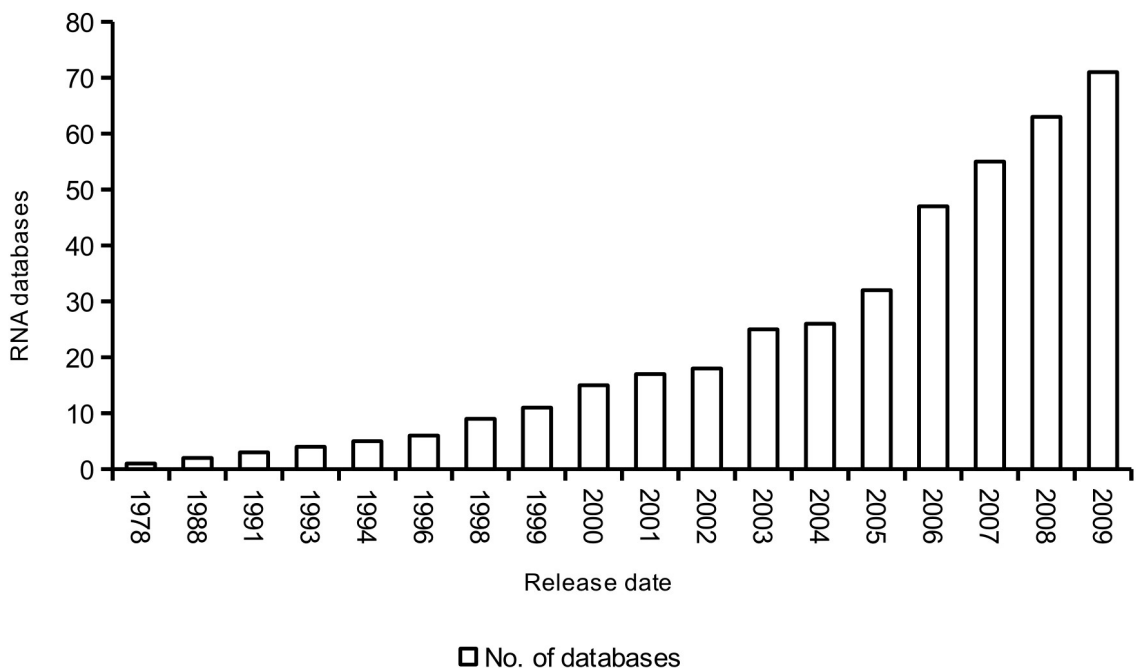

At the time of writing, there are at least 72 active databases dedicated to RNA (9) (http://www.oxfodjournals.org/nar/database/a). Many ncRNAs have inspired their own specialist databases as their functional importance has grown clear. Examples include bacterial small RNAs (10) and eukaryote microRNAs (11), both of which function in the regulation of gene expression. Other abundant classes are small interfering RNAs (siRNA) (12) and small nucleolar RNAs (13). The latter have been found in both archaea and eukaryotes and have spawned several databases over the years, focused on the inventory of single species or on collecting information from a broad taxonomic range

In contrast to these 'specialist' databases and as a consequence of the widening spectrum of known RNA classes, the need for 'generalist databases' has emerged. Examples include the noncoding sequence database Noncode (14) and the RNA family database Rfam (15). Focusing on information relevant only to a narrow functional class of RNA, these databases aim to provide broad information about all RNAs.

The RNA community continues to drive innovation in database design. The Rfam database has recently shifted its annotation to the open encyclopedia, Wikipedia (16). Through Wikipedia anyone with an internet connection can contribute or correct annotation, allowing database developers to focus on adding and improving data sources. Early fears about vandalism seem to be unfounded (15), and the open community annotation model is being adopted by other databases and scientific institutions.

## *Curation and Data*

Given the diversity of RNA function, it is not surprising that different classes of ncRNAs can require different approaches for data production (1). One of the more important challenges is the detection and classification of RNAs. The various classes of RNA have different defining features that need to be taken into consideration. An example of this are microRNAs (miRNAs). Given the relatively short sequence of a mature miRNA (~22 bp), testing for the presence of their characteristic stem-loop structure is an essential step in the prediction process to decrease the chance of false annotations. In contrast, small nucleolar RNAs possess well-conserved sequence motifs in addition to structural features. Similar requirements exist for many other classes as well. Consequently, detection algorithm choice depends on the class of RNA under investigation.

Another distinguishing factor across databases is scope. Some databases focus on a particular type of ncRNA (specialist databases), while others provide access to sequences from a broad range of ncRNA classes (generalist databases). Similarly, sequences may be presented individually (sequence databases) or grouped based on common function and inferred ancestry (alignment databases). Some databases rely on data manually curated by domain experts in addition to or instead of computational predictions. In the following, we will discuss the various approaches used in RNA databases, then present some specific examples that demonstrate their application.

**Manual versus automated annotation**

One of the main differences between RNA databases is whether the data is the result of experimental discovery and verification or derived from automatic, computational scans (discussed in other chapters). The former is often used in specialist databases, whereas the latter generally finds application in genome-scale annotation processes.

Automated annotation has a clear advantage in that it can be used to quickly identify potential RNAs in very large data sets. Relevant tools range from sequence similarity search methods such as BLAST (17,18) to complex probabilistic models (19). Rather then relying entirely on experimentally confirmed results, automated annotation uses some criterion of sequence and/or structural similarity to define thresholds for the inclusion of new sequences (e.g. covariance models as implemented in Infernal (20)). This "thresholding" allows for greater transparency and makes it possible for researchers to apply the same method to their own data. Likewise, new insights into the sequence or structure of a given RNA can be easily incorporated into the annotation process without requiring a time-consuming manual re-evaluation of all data.

Despite technological advances, computational predictions still come with a number of caveats. First, they are only as good as the information they are built on - such as a seed alignment for covariance models. While manual curation of data is potentially subjective and can result in the occasional false annotation, any error in the automated annotation process will affect all down-stream predictions and could be more severe. However, these mistakes can also be more easily rectified by adjusting the relevant parameters and rerunning the analysis.

Second, the degree of sequence divergence of ncRNAs remains a complicated issues and can be highly variable across classes and lineages. Some ncRNAs, such as tRNAs, have relatively well conserved sequences and have been predicted with a low error rate across all domains of life (21). However, many other ncRNAs, such as small nucleolar RNAs (snoRNAs), can exhibit a high degree of sequence plasticity (22). SnoRNAs are involved in the guidance of modifications on other RNAs (mostly rRNA) through the interaction with a conserved protein complex. As their function is determined by their secondary structure and a small stretch of complementarity to their target sequence, primary sequence information may not be sufficient for reliable identification over larger evolutionary distances. In such cases, finding genuine RNA genes without producing many false positives is challenging and experimental validation is of crucial importance.

Whether to prefer manual or automatic curation depends on the intended use of the data. Manual curation will provide high specificity, while automatic curation will produce a higher false positive rate in exchange for a faster and more systematic detection. It should be noted that both approaches can be complimentary. Manual curation is usually the first step in automatic annotation pipelines, while automatic annotation in turn recovers many candidate sequences later subjected to manual inspection and curation. In any case, information from databases can not replace a solid understanding of the biology of the ncRNAs under investigation and scrutiny of the information is always advisable.

**Sequence versus alignment databases**

A large fraction of RNA databases can be further divided into sequence databases and alignment databases. Sequence databases, such as GenBank (23), are primarily designed as repositories. Their primary purpose is to store individual sequences, usually complemented by cross-references to publications and other databases. This approach makes it easier for both developers and external contributors to add to the database, as little specialized analysis beyond sequence discovery is necessary for a new entry.

However, sequence databases generally contain no detailed information on individual ncRNAs, their relationships to each other or a robust nomenclature. The lack of explicit homology information or a stringent nomenclature may negatively impact the value of such resources to certain users. For example, discovering all of the U3 snoRNA sequences in GenBank presents a considerable challenge. Some of the relevant entries may contain helpful information in their descriptions, but many others will not. Even more are probably not annotated at all.

In contrast, to better capture the diversity of a ncRNA, some of the more specialized projects (e.g. miRBase) bin ncRNAs into families based on their expected common ancestry using similarity in structure and sequence. These sequences can then be aligned, producing an estimate of the diversity of each nucleotide. The alignments may subsequently be employed to e.g. train covariance models and thus greatly improving out ability to identify homologs across genomes. On the downside, this process is time consuming as each sequence and alignment need to be manually curated to ensure optimal sensitivity. As such, expansion of these databases may be limited by the availability of expert curators able to perform such tasks.

**[Figure 6.3]**

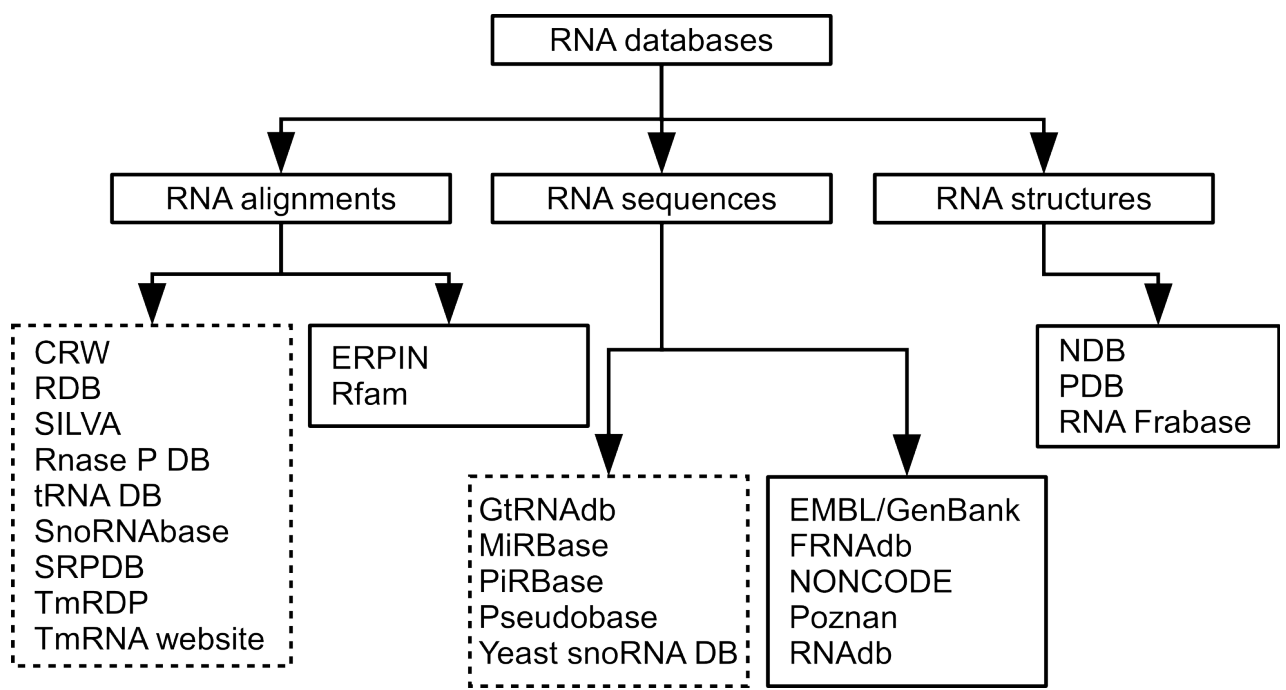

## *RNA databases - examples and practical application*

In the previous section, we defined RNA databases in terms of their curation style and data types. These differences have implications for when and how databases should be used. In the following, we will explore these implications by the example of a few leading databases covering the breadth of methods and scopes.

**General sequence databases - Noncode**

Noncode, established in 2005, is among the most exhaustive general databases on the topic of ncRNAs (14). At the time of writing, it holds information on 112 distinct classes from over 800 species. Noncode is a general sequence database and as such focuses on presenting individual sequences with relevant metadata. In contrast to many other databases, candidate genes are derived primarily from experimental data – over 80% of sequences. This is an impressive feat when considering that there are over 200000 sequence entries in Noncode (release 2.0, 2007).

Data production in Noncode is a step-wise, semi-automated process. Initially, a set of broad keywords is used to identify putative ncRNAs in the literature and new queries are iteratively added to the original list as they are recovered during this search. Based on the results of this process, the GenBank database is then automatically filtered for candidate entries. All data is manually vetted by reference to relevant publications to ensure that they constitute genuine RNAs and to gather additional information relevant to their biological role. Sequences are checked for redundancy before being added to the final data set (Figure 6.4).

[Figure 6.4]

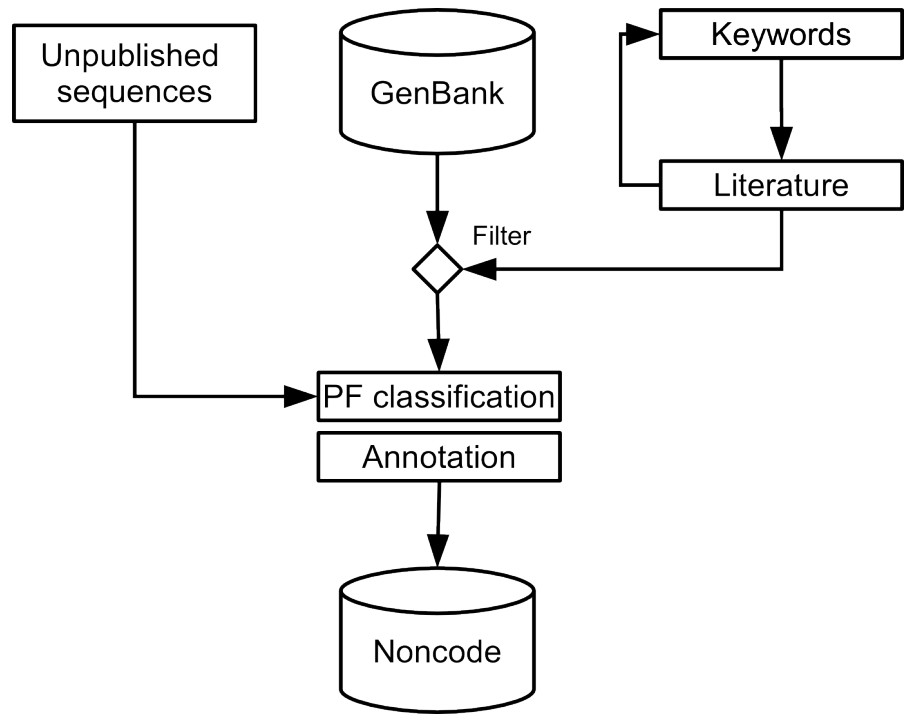

A useful innovation in Noncode is the introduction of a 'process function classification' (PFC). PFC is a vocabulary that describe the cellular functions a ncRNA takes part in, similar to GO terms (24). This system was introduced in an attempt to systematize RNAs functional nomenclature and allows quick access to particular functional classes. Other useful parameters used to describe RNAs include the molecular mechanism of their function, their sub-cellular location, or their cellular role. Noncode can also report ncRNAs based on their organismal range.

Noncode features a boolean search engine that allows users to perform complex queries against the data set, enabling efficient data mining - with certain limitations. Among these, it relies on existing annotation and so does not contain unannotated homologous sequences, though a BLAST sequence search is available for specific sequence queries. Additionally, while all sequences are available as a bulk FASTA-format download, the metadata is not. This limits the prospects for preforming bespoke analysis of the Noncode dataset.

At the time of writing (March 2011) updates to Noncode have been sparse (2005 and 2007), and so may not reflect the current state of ncRNA research. A contributing factor here may be the presumably time-consuming manual vetting of data as part of the production pipeline.

**Specialized sequence databases - miRBase**

The microRNA (miRNA) database miRBase was first released in 2002 as the 'microRNA Registry' (25). It is currently the most complete resource for information on miRNAs, a diverse group of eukaryote RNAs involved in the regulation of gene expression (26). The primary goal of miRBase is to collect published, experimentally verified miRNA sequences and provide researchers with a consistent nomenclature. As of March 2011, the database has seen 16 major releases and contains entries for over 17,000 distinct mature miRNA and their sequences in over 140 species (Figure 6.5).

[Figure 6.5]

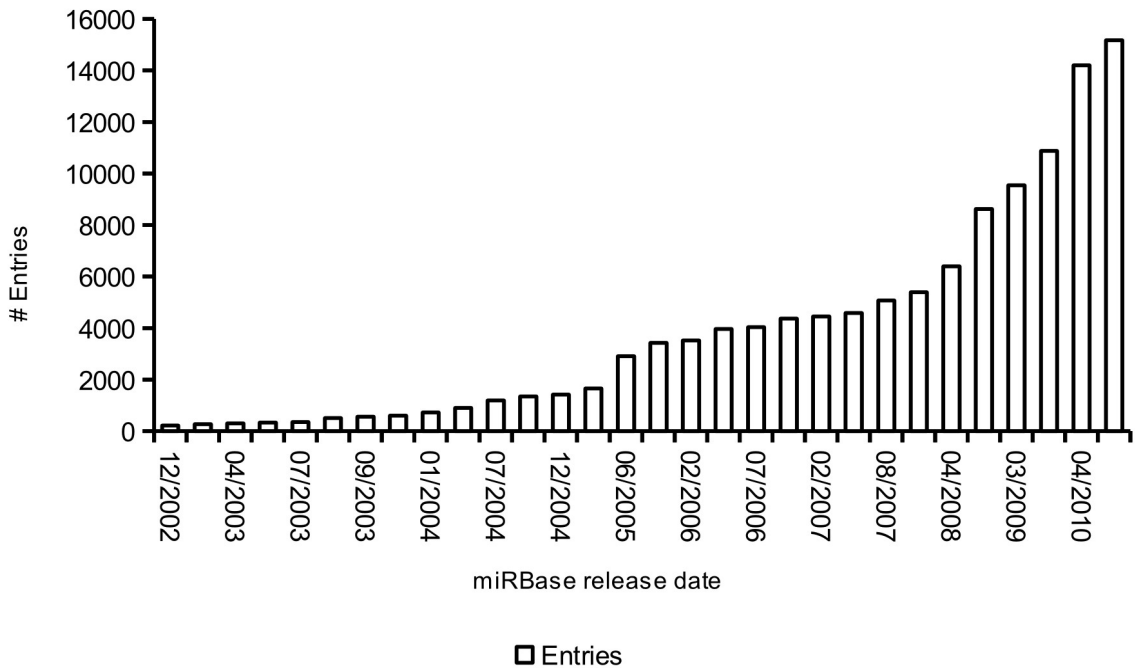

Conceptually, miRBase can be divided into three parts (Figure 6.6). The first is the registry, in which novel miRNAs are included and given a unique id and stable accession number. In order to be considered for inclusion into miRBase, sequences must conform to a set of quality criteria and be either derived from experimental studies or show clear homology to existing entries (27). De-novo computational predictions are not part of the miRBase data set. Submitted sequences are manually inspected and integrated into the database, but only after the work describing the new data has been accepted for publication in a peer-reviewed journal. This serves to maintain a high standard and clean data set with few false annotations. Where applicable, entries are also assigned to a higher order family which group miRNAs across species based on common ancestry. Owing to the increasing amount of available sequencing data, miRBase has recently integrated new procedures to recover miRNAs from the Gene Expression Omnibus database (28) using, among other things, significant sequence similarity to existing entries and characteristic expression profiles as criteria to identify genuine miRNAs

The second part of miRBase focuses on miRNA sequences. Individual miRNAs are annotated in depth with information such as genome coordinates and genomic context (intronic, intergenic, clustered or singleton). Relevant meta data is extracted from the literature, including details on experimental procedures used to identify and characterize the respective genes. Finally, miRBase links individual miRNA entries to external databases that focus on automated prediction of possible mRNA targets, such as 'microCosm' which was originally developed as part of miRBase.

[Figure 6.6]

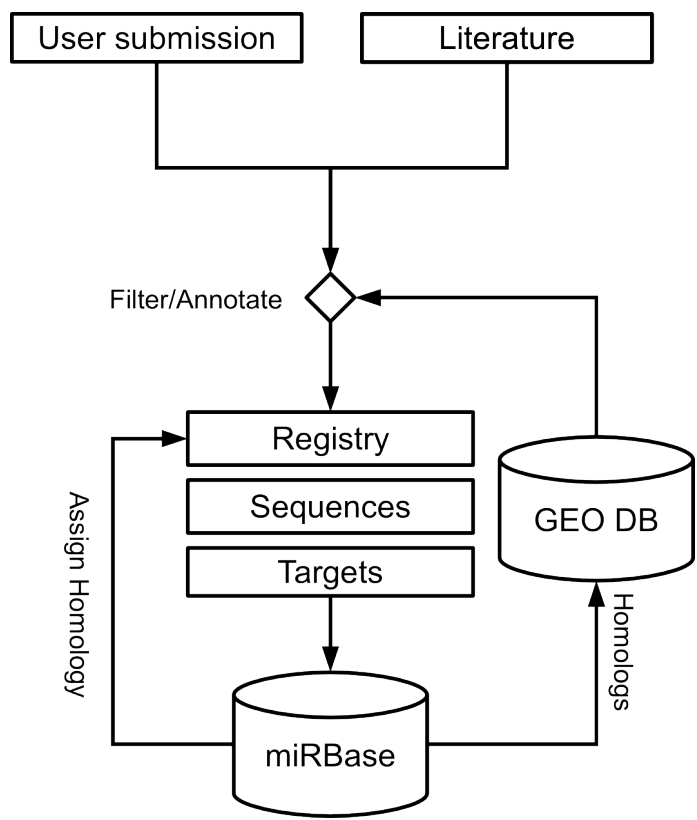

MiRBase offers a range of options for users to interact with its data. Most importantly, it can be searched for a set of pre-defined criteria including species, genomic location and expression patterns. While a free-form search is possible, it does not support complex boolean queries. In addition, BLASTN and SSEARCH can be used to find known miRNAs similar to a query sequence. The wealth of metadata makes miRBase particularly useful for bioinformaticians. Not only is it possible to download all the sequences, but the entire database is freely available as well. This opens up the possibility of designing custom data mining pipelines which incorporate all of the information contained in miRBase.

**General alignment databases - Rfam**

The RNA family database, Rfam, is the largest general alignment database currently available (15,29). As of release 10, it provides over 3 million annotations for 1446 distinct ncRNAs across the entire EMBL nucleotide database. Rfam defines all sequences that align to a covariance mode constructed from a "seed" alignment within certain sequence and structural similarity criteria as a family. These "seed" models are constructed from two or more representative sequences and are manually curated, using published data. The thresholds of Rfam covariance models are individually adjusted to account for the varying degree of sequence plasticity specific to each family, reducing the fraction of false annotations. A possible shortcoming of this approach is that distantly related RNAs may not be captured by a model. To address this issue, Rfam has introduced clans, which are a higher-order grouping of families based on expected common descent using information from the literature and measures of similarity between families (16). Clans and families provide a robust nomenclature to identify homologous RNAs across genomes.

[Figure 6.7]

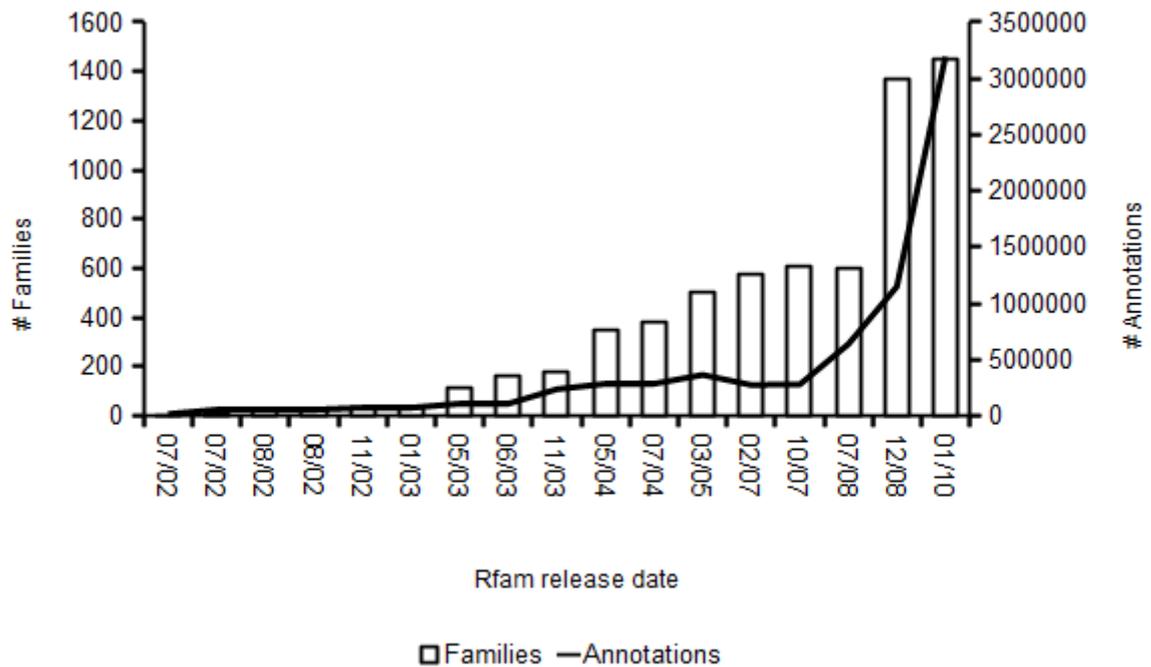

The Rfam annotation pipeline consist of a two step process (Figure 6.8). First, verified RNAs from the literature or external databases are used to create seed alignments. Sequences from these families are then used as queries for WU-BLAST searches against the EMBL nucleotide database to identify candidate RNAs. This step is made necessary by the comparatively high computational requirements of the covariance model search but may be replaced by accelerated profile hidden Markov models in the near future (30,31). All candidate sequences are subsequently subjected to more rigorous covariance model searches (20). These models are calibrated so as to capture the suspected range of an ncRNA family, while providing a low false positive rate.

**[Figure 6.8]**

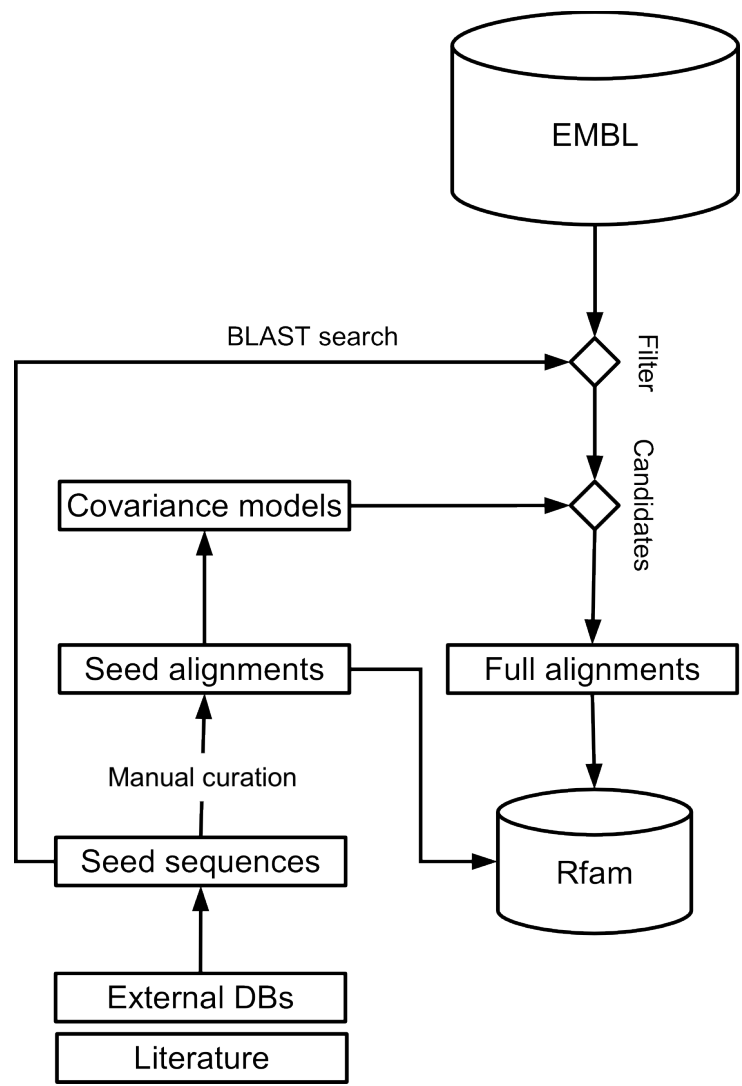

The tools and information provided by Rfam are diverse. Full alignments of all annotations as well as evolutionary trees are available to provide users with insight into the diversity and phylogenetic distribution of each family. In addition, users can search sequences against the Rfam database to identify putative ncRNAs. In addition, the Rfam database as well as the annotation pipeline are available for download from the website. Rfam has also found application in, for example, the Ensembl database (32) where it is used to provide annotations of ncRNAs in a range of different genomes. Rfam annotations are provided through Wikipedia, allowing researchers to rapidly update and correct annotations as new information becomes available.

The use of computational predictions means that Rfam provides ncRNA annotations for many genomes that have not yet been studied in detail. However, there are limitations to this approach. Despite rigorous examination and thresholding of families, the Rfam pipeline can produce false positives. Secondly, Rfam is not exhaustive but features only a limited number of ncRNAs. Newly discovered ncRNAs need to be manually curated before being included in the database, which requires time. In an effort to speed up this process, Rfam has introduced a special publication track in collaboration with the journal RNA Biology where researchers can publish alignments and Wikipedia annotations of the ncRNAs they work on, which are then included in the database.

**Specialized alignment databases - SILVA**

Ribosomal RNAs (rRNAs) were among the first RNAs to be cataloged in databases, and several projects have done this over the years. This continued interest is partially due to the fact that rRNAs are widely used as phylogenetic markers. This has special relevance to the emerging field of metagenomics, where the taxonomic diversity of a sample can be difficult to determine.

At the time of writing, the most complete rRNA database is SILVA, which was originally released in 2007 (33). The release 104 (October 2010) holds information on almost 1.5 million small subunit (SSU) and over 200.000 large subunit (LSU) rRNA entries from all domains of life. The SILVA release cycle is synchronized with the EMBL nucleotide database, from which it draws sequence data. As a result, it is updated on a regular basis.

**[Figure 6.9]**

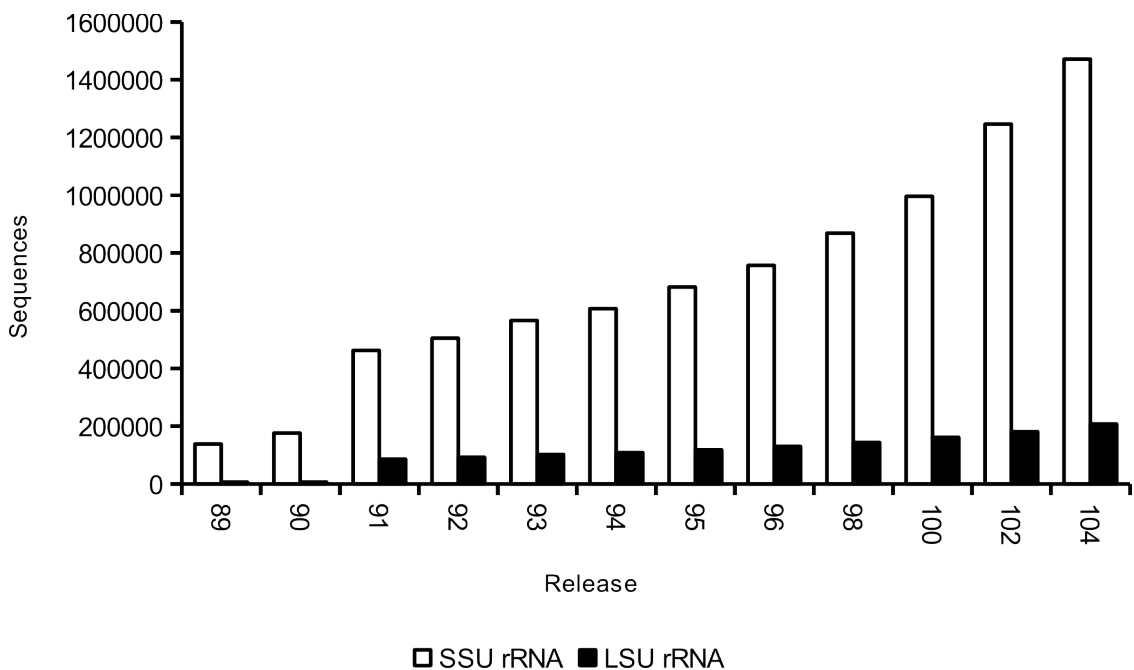

SILVA's automated production pipeline is based on a set of keywords that are used to identify putative rRNAs in the EMBL database, given existing annotations or descriptions. To account for the increasing number of unannotated rRNAs produced by large-scale sequencing projects, EMBL sequences are scanned using Hidden Markov Models to identify additional candidates (34,35). Data retrieved in these ways are subsequently filtered based on a set of stringent criteria aimed at identifying genuine rRNA genes, including a minimum size requirement and a maximum number of allowed ambiguous positions. All rRNAs which pass these filters are then aligned against aforementioned seed alignments and stored in the database. In addition to this primary, comprehensive data set (referred to as 'Parc'), SILVA compiles the 'Ref' data set, a subset of 'Parc' comprised of high-quality, full- or nearly full-length sequences.

**[Figure 6.10]**

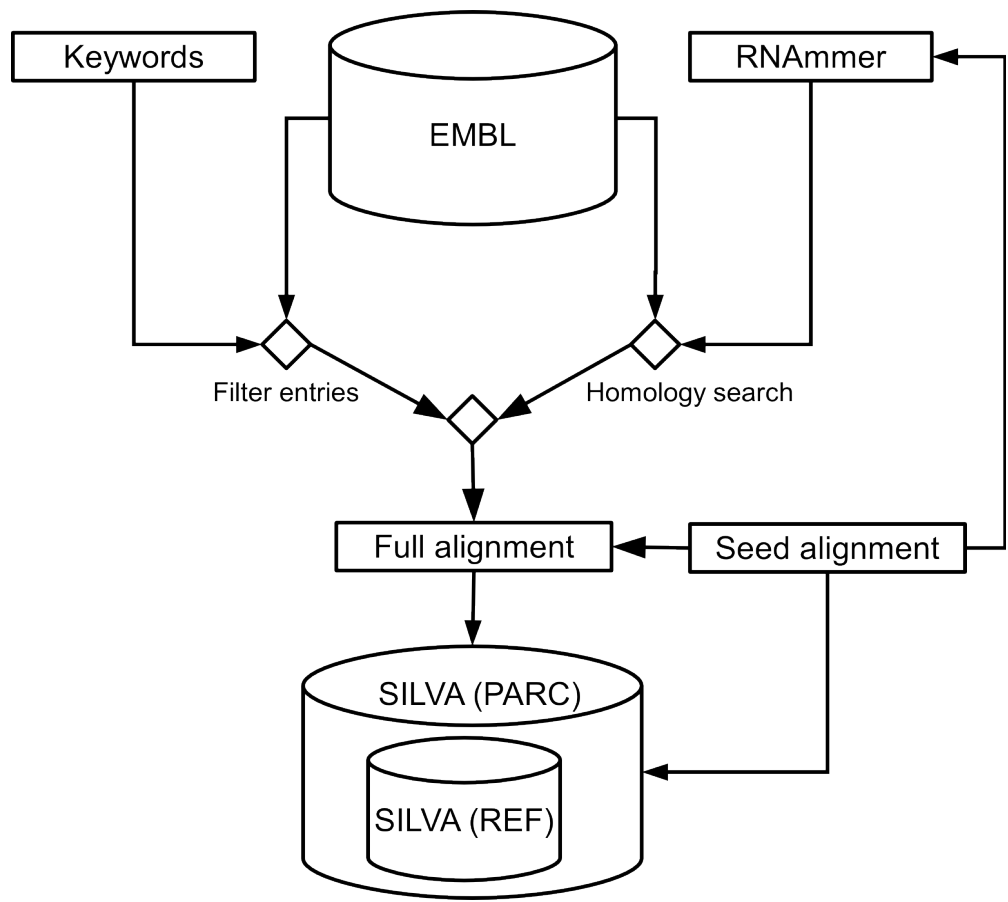

SILVA provides users with a range of tools. Most importantly, annotations can be searched using numerous criteria, including organism names, accession numbers or related publications. All retrieved hits can be downloaded in either FASTA or the ARB format. Another helpful feature is the use of ontologies, such as the environmental or the aforementioned taxonomic affiliation, to further characterize sequences. Finally, user-submitted queries can be aligned against the respective seed alignments, or subsets thereof, to create quality alignments from their own data .

Data from SILVA is free for academic use and pre-filtered datasets are available for download.

**Summary**

[Table 6.1 near here]

## *Closing remarks*

Two major trends have been driving the development of ncRNA databases: an increasing appreciation of the importance of ncRNA genes, particularly in the face of high-throughput sequencing technologies, and the development of faster, more accurate computational tools for identifying ncRNA sequences. We have presented four databases here with very different approaches to making sense of increasingly large data sets. We believe these approaches are complimentary, as these diverse molecules demand diverse approaches to their characterization. It is important that users be aware of the potential for false positives, particularly in computationally-produced predictions. In these cases, cross-validation from multiple sources can provide higher certainty in an annotation.

## *Figure captions*

**Figure 6.1: The expanding RNA world.** The expanding picture of non-coding RNAs as established by the number of publications dedicated to the subject matter and listed in the PubMed database (identified by the tag 'npcRNA').

**Figure 6.2: Number of RNA databases.** The growth in available data and general interest in the various classes of RNAs is mirrored in the increase of the number of RNA databases (based on data from http://www.oxfordjournals.org/nar/database/a/).

**Figure 6.3:** RNA databases can be broadly organized into alignment, sequence and structure databases. These are further grouped into class specific databases (dotted line) and general databases (solid line). An index of RNA databases is maintained at http://www.oxfodjournals.org/nar/database/a.

**Figure 6.4:** The Noncode annotation pipeline compiles experimentally verified ncRNAs from the literature and the nucleotide database GenBank using a set of relevant keywords. In addition, the authors also include unpublished RNAs studied in their lab. The resulting data is subjected to manual inspection and annotation (such as the Noncode-specific process function classification) prior to inclusion in the database.

**Figure 6.5:** The miRBase database was first released in 2002 and as of early 2011 has seen 16 major releases. During this time it has grown significantly, mirroring closely the growth of the RNA field and the increasing amount of sequence data.

**Figure 6.6:** Data in miRBase stems primarily from experimental evidence. After submission, new miRNAs are given an entry in the registry, annotated and linked to external sources for target prediction. In an attempt to capitalize on the increasing amount of sequencing data, miRBase more recently expanded its pipeline to identify expressed miRNA candidates in the GEO database, based on expected homology to existing entries.

**Figure 6.7:** Release 10.0 of Rfam contains information on 1446 ncRNA families, yielding more than 3 million annotations across the EMBL nucleotide database. Each family is defined as sequences aligning to a co-variance model, based on manually curated seed alignments from published ncRNA data.

**Figure 6.8: The Rfam pipeline.** The Rfam pipeline consists of two major steps. First, experimentally verified sequences are grouped based on their expected common ancestry to create seed alignments. Sequences from these alignments are then used in a traditional BLAST search against the EMBL database to identify putative ncRNA genes. To further increase the confidence in these predictions, all candidate hits are analyzed with manually curated co-variance models. Sequences that pass both these tests are included in Rfam.

**Figure 6.9:** Since its initial release in February 2007 (89, based on EMBL 89), data for both small subunit rRNA (SSU) as well as large subunit rRNA (LSU) has increased markedly over the course of only 3 years. As of release 104, SILVA contains information on over 1.4 million SSU sequences and more than 200.000 LSU sequences.

**Figure 6.10: SILVA data production.** The SILVA rRNA database is built in three automated steps. Starting with a set of keywords, putative ribosomal RNAs are retrieved from the EMBL nucleotide database. rRNA candidates are identifed by alignment to a curated profile hidden Markov model. The resulting data then has to meet a number of criteria, including a minimum length of 300 bases and a maximum of 2% ambiguities, for inclusion in the database. The comprehensive dataset of SSU and LSU sequences is referred to as 'Parc', from which a subset of full-length, high quality sequences is created ('Ref').

*Tables*

**Table 6.1:** Summary and comparison of major RNA databases

|  | **Noncode** | **Rfam** | **miRBase** | **SILVA** |
|---|---|---|---|---|
| DB type | General sequence DB | General Alignment DB | Specialist sequence DB | Specialist Alignment DB |
| Manual annotation | yes (based on published data) | yes (seed alignments and Wikipedia) | yes | yes (seed alignments) |
| Comp. annotation | no | yes | yes (homologs only) | yes |
| Data source | GenBank/Literature | EMBL | User submission | EMBL |
| Nomenclature | no | yes | yes | NA |
| Data download | sequence files | everything | everything | sequence files |
| Release cycle | irregular | ~1-2/year | ~1-2/year | synced with EMBL |